\documentclass{llncs}
\usepackage[show]{ed}
\usepackage{calbf}
\usepackage{amstext,amsmath,amssymb}
\usepackage{xspace}
\usepackage[T1]{fontenc}
\usepackage{caption}

\usepackage{mdframed}
\newenvironment{boxedquote}{\begin{mdframed}[leftmargin=1cm,rightmargin=1cm]}{\end{mdframed}}

\usepackage{wrapfig,paralist}
\usepackage[hyperref,backend=bibtex,style=alphabetic]{biblatex}
\renewbibmacro*{event+venue+date}{}
\renewbibmacro*{doi+eprint+url}{%
  \iftoggle{bbx:doi}
    {\printfield{doi}\iffieldundef{doi}{}{\clearfield{url}}}
    {}%
  \newunit\newblock
  \iftoggle{bbx:eprint}
    {\usebibmacro{eprint}}
    {}%
  \newunit\newblock
  \iftoggle{bbx:url}
    {\usebibmacro{url+urldate}}
    {}}

\usepackage{tikz}\usetikzlibrary{mmt,fit}

\usepackage[final,today]{svninfo}
\svnInfo $Id: note.tex 143 2014-04-25 16:15:53Z jcarette $
\svnKeyword $HeadURL:s https://svn.omdoc.org/repos/omdoc/doc/epc/paper.tex $

\def\defeq{:=}
\newcommand{\seq}[1]{{\left[ #1 \right]}}
\newcommand{\set}[1]{\left\{ #1 \right\}}
\newcommand{\wf}[2]{{#1\vdash #2~\textbf{wf}}}
\newcommand{\thycons}[2]{{#1\ltimes #2}}
\newcommand{\redge}[1]{\mathrel{\xrightarrow{#1}}}

\newcommand{\view}[3]{\ensuremath{{#1}:{#2}\rightsquigarrow{#3}}}
\newcommand{\UDLM}{\textsf{UDLM}\xspace}
\def\defemph#1{\textbf{#1}}
\def\ul#1{\underline{#1}}

\newcommand{\nat}{\ensuremath{\mathbb{N}}\xspace}
\newcommand{\natprime}{\ensuremath{\mathbb{N}'}\xspace}
\newcommand{\reals}{\ensuremath{\mathbb{R}}\xspace}
\newcommand{\realsprime}{\ensuremath{\mathbb{R}'}\xspace}
\newcommand{\sfour}{\textbf{S4}\xspace}

\usepackage{hyperref}
\title{Realms: A Structure for Consolidating Knowledge about Mathematical Theories\thanks{Work partially supported by NSERC.  Final publication
is available at \url{http://link.springer.com}}}
\author{Jacques Carette\inst{1} \and William M. Farmer\inst{1} \and 
Michael Kohlhase\inst{2}}
\institute{
  Computing and Software, McMaster University\\
\url{http://www.cas.mcmaster.ca/~carette}\\
\url{http://imps.mcmaster.ca/wmfarmer}
\and
  Computer Science,  Jacobs University Bremen\\
  \url{http://kwarc.info/kohlhase}
}

\begin{document}
\maketitle
\begin{abstract}
Since there are different ways of axiomatizing and developing a
mathematical theory, knowledge about a such a theory may reside in
many places and in many forms within a library of formalized
mathematics.  We introduce the notion of a \emph{realm} as a structure
for consolidating knowledge about a mathematical theory.  A realm
contains several axiomatizations of a theory that are separately
developed.  Views interconnect these developments and establish that
the axiomatizations are equivalent in the sense of being mutually
interpretable.  A realm also contains an external interface that is
convenient for users of the library who want to apply the concepts and
facts of the theory without delving into the details of how the
concepts and facts were developed.  We illustrate the utility of
realms through a series of examples.  We also give an outline of the
mechanisms that are needed to create and maintain realms.

\end{abstract}

\section{Introduction}\label{sec:intro}

In \cite{Farmer:mkm11} the second author calls for the establishment
of a ``universal digital library of mathematics'' (\UDLM). In our
joint work it is understood that the {\UDLM} will be organized as a
theory graph, i.e., a set of theories (collections of symbol
declarations, definitions, assertions, and their proofs)
interconnected by meaning-preserving views (morphisms). The
``little/tiny theories approach'' first put forward
in~\cite{FaGu:lt92} has been very fruitful for formal developments of
mathematical knowledge, but it has not found its way into mainstream
mathematics. One reason may be that there is a mismatch with the way
mathematicians --- the supposed users of the {\UDLM} --- think about
and work with theories. \cite{CarFar:hlt08} argues for the development
of ``high-level theories'' that better mesh with these
expectations. We will re-examine the issues involved and
propose a solution.

In the mathematical community the term ``\emph{theory}'' is used to
describe multiple ideas, from the axiomatic theory of the algebraic
structure of a group to ``Group Theory'' as an entire discipline, and
various gradations in between.  Looked at more closely, this implies a
multi-scale organizational structure to the basic components of
mathematics, ranging from individual concepts (e.g., a group) to whole
subareas of mathematics (e.g., Group Theory).  Here our interest in
this structure is purely pragmatic: how can it be leveraged to build a
better mechanized mathematics systems and, ultimately, a better
{\UDLM}.

We will consider this in a bottom-up manner: what is the most natural structure on
\emph{theories}
that allows us to abstract away from irrelevant details, yet still allow us to get some
practical work done?  One such structure is that of \emph{mutual interpretability} between
theories.  Basically this is the case when we have two equivalent theories $T_1$ and $T_2$
(in a sense to be made precise later) with presentations that can be markedly different.

But why should \emph{theory presentations} matter at all?  Studies of ``theories'' in
mathematics (e.g., Lawvere theories~\cite{LawvereFW:funsattac,NotionsOfLawvereTheory}) or
in logic focus on entities that are \emph{complete} in some sense.  But such completeness
generally also implies that the object at hand is effectively infinite, and thus cannot be
directly represented in software.  Hence we are immediately forced to work with
\emph{finite representations} of these infinite objects.  Furthermore, by G\"{o}del's
incompleteness theorem, most of the interesting theories will be fundamentally incomplete,
in that no finite representation will be able to adequately represent the complete whole.
The relevance here is that we are forced to deal with (syntactic) representations, which
will generally be \emph{incomplete}.  As this is forced on us, we need to gracefully adapt
to dealing with theory presentations in place of the theories they represent.

\section{The Setting: Theory Graphs}\label{sec:thy-grph}

We will now present an abstract notion of a theory graph that is sufficient to introduce
the notion of a realm without committing ourselves to a particular approach such
as~\cite{CarCon:tpc12} or~\cite{RabKoh:WSMSML13}.

Let a \defemph{theory} be a presentation of an axiomatic theory consisting of a finite
sequence of symbol and formula declarations.  The symbols denote concepts and the formulas
denote facts about these concepts.  There are three kinds of formula declarations:
\defemph{axioms}, \defemph{definitions}, and \defemph{theorems}.\footnote{If we make use
  of the Curry-Howard isomorphism --- as we do in~\cite{RabKoh:WSMSML13}, then we can get
  by with typed symbol declarations (with optional definitions) only. In the
  propositions-as-types paradigm, axioms are typed constant declarations, and theorems are
  typed definitions --- a proof corresponds to the respective definiens of a symbol which
  is of the respective type.} We further assume that for a theory $T =
\seq{A_0,A_1,\ldots,A_n}$, for all $i$ with $0 \le i < n$, $A_{i+1}$ is well formed in the
context of $\seq{A_0,A_1,\ldots,A_i}$.  A theory thus represents an axiomatization of a
mathematical topic.  If $T$ is a theory and $A$ is a symbol or formula declaration, $\wf T
A$, means that $A$ is well formed in the context of theory $T$.  When $\wf T A$, we define
$\thycons T A$ to mean $\seq{A_0,A_1,\ldots,A_n,A}$; we also extend $\ltimes$ to apply to
sequences of declarations (on the right).  If $T$ is a theory and $\varphi$ is a formula,
then $T \models \varphi$ means $\varphi$ is a logical consequence of $T$.  A theory is
\defemph{primitive} if it contains only symbol declarations and axioms.  A primitive
theory represents a set of concepts and facts without a development structure. The
\defemph{empty theory} is the empty sequence.  When $T_1 = \thycons T A$ and $T_2 =
\thycons T B$, we also define $T_1 \oplus T_2 \defeq \thycons {(\thycons T A)} B$ whenever
$A$ and $B$ are disjoint (i.e., they declare symbols or formulas with different names).
By also extending $\oplus$ to sequences of declarations, we get a \defemph{join} operation
on theory presentations.

A \defemph{theory graph} is a directed graph in which a node is a theory and we have edges
from a theory $T$ to a theory $\thycons T A$.  If $T$ and $T'$ are theories in a theory
graph $G$, an edge from $T$ to $T'$ is designated as $T \redge{G} T'$.  A theory graph is
a modular representation of a formalized body of mathematical knowledge.

An \defemph{axiomatic development} of a theory $T$ to a theory $T'$ in
$G$ is a subgraph $G'$ of $G$ in which $T$ is the only source of $G'$
and $T'$ is the only sink of $G'$.  In this case, $T$ and $T'$ are
called the \defemph{bottom theory} and the \defemph{top theory},
respectively.  An axiomatic development is thus a lattice of theories
in which the top theory is the join of the members of the lattice.

Let $T_1$ and $T_2$ be theories in $G$.  A \defemph{view of $T_1$ in
  $T_2$} is a homomorphic mapping $\Phi$ of the language of $T_1$ to
the language of $T_2$ such that, for each formula $\varphi$ of $T_1$,
$T_1 \models \varphi$ implies $T_2 \models \Phi(\varphi)$.  We denote
a view by $\view \Phi {T_1} {T_2}$.  A view is thus a meaning
preserving mapping that shows how $T_1$ can be embedded in $T_2$.  It
also provides a mapping from the models of $T_2$ to the models of
$T_1$ (note the reversal of order).  $T_1$ and $T_2$ are
\defemph{equivalent} if there is a view in both directions.  A view
$\view \Phi {T_1} {T_2}$ is \defemph{faithful} if for each formula
$\varphi$ of $T_1$, $T_2 \models \Phi(\varphi)$ implies $T_1 \models
\varphi$.  Views give a second (oriented, multi) graph structure on
$G$, making it into a bigraph.  It is important to note that the base
theory graph (with edges but not views) is always acyclic.

Let $T$ and $T'$ be theories.  $T'$ is an \defemph{extension} of $T$, and $T$ is a
\defemph{subtheory} of $T'$, if there exists a sequence $S$ such that $T' = \thycons T S$.
In this case, there is a view $\view \Phi T {T'}$ such that $\Phi$ is the identity
function. We call $\Phi$ the \defemph{inclusion} of $T$ into $T'$ and denote it by $T
\mmtar{include} T'$.  This corresponds to the extensions of~\cite{CarCon:tpc12}, identity
structures in~\cite{RabKoh:WSMSML13}, as well as the display maps of categorical type
theory.

An \defemph{interface} for $T$ is a view $\view \Phi {T'} T$ such that $\Phi$
is injective.  $T'$ and $T$ are called, respectively, the \defemph{front} and
\defemph{back} of the interface.  Each subtheory of $T$ can be a front of an interface for
$T$.  An interface is intended to be a convenient means for accessing (parts of)
$T$.  The front of a good interface includes a carefully selected set of symbols and
formulas that denote orthogonal concepts and facts that can be easily combined to express
the other concepts and facts of $T$.  See section~\ref{sec:examples} for some concrete
examples.

\begin{wrapfigure}r{2.5cm}\vspace*{-2em}
\begin{tikzpicture}[xscale=1.6,yscale=1.6]
\node[thy] (at) at (0,0) {$T$};
\node[thy] (atp) at (0,1) {$T'$};
\draw[conservative] (at) -- (atp);

\node at (.4,.5) {$\defeq$};

\node[thy] (t) at (1,0) {$T$};
\node[thy] (tp) at (1,1) {$T'$};
\draw[include] (t) to [out=70,in=-70] node[right]{$\iota$}(tp);
\draw[view] (tp) to [out=-110,in=110] node[left]{$\Phi$} (t);

\end{tikzpicture}\vspace*{-2em}
\end{wrapfigure}
An extension $T'$ of $T$ is \defemph{conservative} if there is a view $\view \Phi {T'} T$
such that $\Phi$ is the identity function when restricted to $T$ (i.e.,
$\Phi\circ\iota=\text{Id}_T$ where $\iota$ is the inclusion of $T$ in $T'$).  If $T'$ is a
conservative extension of $T$, then for each formula $\varphi$ of $T$, $T' \models
\varphi$ implies $T \models \varphi$. Common examples of conservative extensions are
extensions by symbol declarations, definitions, or theorems (with proofs).  Obviously, if
$T'$ is a conservative extension of $T$, then $T$ and $T'$ are equivalent.  
We abbreviate the two arrows in a conservative extension with a double inclusion arrow in
theory graphs.  A subgraph $G'$ of a theory graph $G$ is conservative if $T'$ is a
conservative extension of $T$ for each edge $T \redge{G} T'$ in $G'$.  A
\defemph{conservative development} is an axiomatic development that is conservative.  Note
that all the theories in a conservative development are equivalent to each other. We will
write a conservative development with bottom theory $S$ and top theory $T$ as
$S\mmtar[xscale=1.5]{conservdev}T$.

$\view \Phi {T_1} {T_2}$ is \defemph{expansive} if there is a $\view \Psi {T_2} {T_2}$
such that (1) the range of $\Psi$ is the image of $\Phi$ and (2) $\Psi$ is the identity on
the image of $\Phi$.  That is, a view of $T_1$ in $T_2$ is expansive if, roughly speaking,
$T_2$ is a conservative extension of the view's image.  A view of $T$ is
\defemph{conservative} if it is faithful and expansive.  A conservative view of $T$ is a
generalization of a conservative extension of $T$.  If there is a conservative view of
$T_1$ in $T_2$, then $T_1$ and $T_2$ are mutually viewable.

\section{Motivation: Developers, Students, and Practitioners}\label{sec:motivation}

The user of a {\UDLM} can play three different roles.  As a \emph{developer}, the user
creates new representations of mathematical knowledge or modifies existing ones in the
library.  As a \emph{student}, the user studies the mathematical knowledge represented in
the library.  And, as a \emph{practitioner}, the user applies the mathematical knowledge
in the library to problems, both theoretical and practical.  A user may play different
roles at different times and may even sometimes combine roles.

A theory graph does not fully support all three of the user's roles.
In fact, it lacks the structure that is necessary to satisfy the
following requirements:


\def\realmref#1{\textbf{R\ref{realmref:#1}}}
\begin{compactenum}[\bf R1]

\medskip

\item\label{realmref:where} There can be many \emph{equivalent theories} in a theory graph
  that represent different axiomatizations of the \emph{same mathematical topic}.  As a
  result, concepts and facts about this mathematical topic, possibly expressed in
  different languages, may be widely distributed across a theory graph.  The developer and
  the student would naturally want to have these \emph{different axiomatic developments}
  and the \emph{set of concepts and facts} that are produced by them in \emph{one
    convenient place and in one convenient language}.

\item\label{realmref:minimality} Developers prefer developments that start with
  \emph{minimal bottom theories} and are built as much as possible using
  \emph{conservative extensions}.  This approach minimizes the chance of \emph{introducing
    inconsistencies} (which would render the developments pointless) and maximizes the
  \emph{opportunities for reusing the development in other contexts}.  While these two
  benefits may not be of primary concern for the student and the practitioner, such a
  careful development is usually easier to understand and produces concepts and facts that
  can be more reliably applied.

\item\label{realmref:views} The developer would like to \emph{create
  a view from one theory to another in a convenient manner} by
  starting with a view of a minimal axiomatization of the theory and
  then building up the view as needed using conservative extensions.
  Also, there is a desire to use \emph{the most convenient axiomatization}
  amongst equivalent presentations.

\item\label{realmref:users1} The \emph{application of a mathematical
  fact} usually does not require an understanding of how concepts and
  facts were derived from first principles.  Hence the parts of the
  theory graph which were needed by the developer may not be useful
  to the practitioner, and may well get in the way of the practitioner's work.
  The practitioner would naturally want the concept or fact to be \emph{lifted
  out of this tangled development bramble}.

\item\label{realmref:users2} Languages are introduced in a theory
  graph for the purpose of theory development.  They may employ
  vocabulary that is inconvenient for particular applications.  The
  practitioner would naturally like to have \emph{vocabulary chosen
    for applications instead of development}.
\end{compactenum}

\medskip

\noindent
In summary, the developer, the student, and the practitioner have
different concerns that are not addressed by the structure of a theory
graph.  These different concerns lead us to propose putting some additional
structure on a theory graph, a notion we call a ``realm'', to meet these
five requirements.

\section{Realms}\label{sec:realms}

In a nutshell, a realm identifies a subgraph of a development graph,
equips it with a carefully chosen interface theory that abstracts from
the development, and supplies the practitioner with the symbols and
formulas she needs.
 
\begin{definition}\label{def:realm}\rm
  A \defemph{realm} $R$ is a tuple $(G,F,\cC,\cV,\cI)$ where:
  \begin{compactenum}
  \item $G$ is a theory graph.
  \item $F$ is a primitive theory in $G$ called the \defemph{face} of the realm $R$.
  \item $\cC$ is a set $\set{C_1,C_2,\ldots,C_n}$ of conservative developments in $G$.
  \item $\cV$ is a set of views that establish that the bottom theories
    $\bot_1,\bot_2,\ldots,\bot_n$ of $C_1,C_2,\ldots,C_n$, respectively, are pairwise
    equivalent.
  \item $\cI$ is a set $\set{I_1,I_2,\ldots,I_n}$ of conservative interfaces such that $F$ is
    the front of $I_i$ and the top theory $\top_i$ of $C_i$ is the back of $I_i$ for each
    $i$ with $1 \le i \le n$.
  \end{compactenum}
  For each $i$ we call $(\bot_i,C_i,\top_i)$ the \defemph{$i$-th pillar} of $R$ and $I_i$
  its \defemph{interface}. Note that every subset of pillars of a realm $R$ forms a realm
  together with its interface and the face of $R$. We call realms with just one pillar
  \defemph{simple} and realms with more than one pillar \defemph{proper}.
\end{definition}

\begin{wrapfigure}r{6.4cm}\vspace*{-2em}
\begin{tikzpicture}[xscale=.9]
  \node[thy] (t1) at (0,1.5) {$\bot_1$};
  \node[thy] (t2) at (2,1.5) {$\bot_2$};
  \node[thy] (tdots) at (4,1.5) {\ldots};
  \node[thy] (tn) at (6,1.5) {$\bot_n$};

  \draw[view] (t1) to[out=15,in=165] (t2);
  \draw[view] (t2) to[out=195,in=-15] (t1);
  \draw[view] (t2) to[out=15,in=165] (tdots);
  \draw[view] (tdots) to[out=195,in=-15] (t2);
  \draw[view] (tdots) to[out=15,in=165] (tn);
  \draw[view] (tn) to[out=195,in=-15] (tdots);

  \node[thy] (ts1) at (0,3) {$\top_1$};
  \node[thy] (ts2) at (2,3) {$\top_2$};
  \node[thy] (tsdots) at (4,3) {\ldots};
  \node[thy] (tsn) at (6,3) {$\top_n$};

  \draw[conservdev] (t1) -- node[right]{$C_1$}(ts1);
  \draw[conservdev] (t2) -- node[right]{$C_2$} (ts2);
  \draw[conservdev] (tdots) -- (tsdots);
  \draw[conservdev] (tn) -- node[left]{$C_n$} (tsn);

  \node[primthy,minimum width=6cm,outer sep=3pt] (t) at (3,4.2) {$F$};
  \draw[mview] (t) -- node[below]{$I_1$} (ts1);
  \draw[mview] (t) -- node[right,near end]{$I_2$} (ts2);
  \draw[mviewleft] (t) -- node[left,near end]{$I_{\ldots}$} (tsdots);
  \draw[mviewleft] (t) -- node[below]{$I_n$} (tsn);
  \draw[thy,dashed,double] (-.5,1.1) rectangle (6.5,4.6);
\end{tikzpicture}
\caption{The Architecture of a Realm}\label{fig:realm}\vspace*{-2em}
\end{wrapfigure}
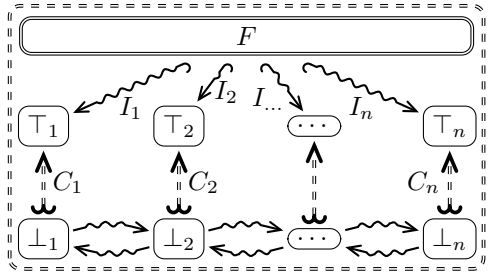
Figure~\ref{fig:realm} shows the general situation, we depict realms by double dashed
boxes and faces by dashed ones.  All the theories in the realm $R$ are equivalent to each
other since
\begin{inparaenum}[\em i\rm)]
\item all the bottom theories are equivalent by the views in $\cV$,
\item all the members of a conservative development are equivalent, and
\item the front and back of a conservative interface are equivalent.
\end{inparaenum}

To ensure that realms have a pleasant categorical structure (that of a contractible
groupoid), we assume that $\cV$ always contains identity views which show
self-equivalence.

A realm consolidates a body of formalized mathematics pertaining to
one topic.  Each bottom theory $\bot_i$ is a different (ideally
minimal) axiomatization and each conservative development $C_i$ is a
family of extensions of $\bot_i$.  $F$ is an (ideally convenient)
presentation of the topic without any development structure and
without any \emph{scaffolding}, i.e., the concepts and facts that are
needed only for development purposes.  Finally, each interface $I_i$
establishes that $F$ is indeed a presentation of the topic and how it
embeds into each $\top_i$.

The realm $R = (G,F,\cC,\cV,\cI)$ minus $F$ and $\cI$ records the
development structure of the topic; we call
$\overline{R}\defeq(G,\cC,\cV)$ the \defemph{body} of $R$.  It can be
used to study the development structure of the topic or as a basis for
extensions.  The face $F$ exposes the most important and useful
concepts and facts pertaining to that realm.  It is also meant to be
used as a module for constructing larger bodies of formalized
mathematics.  In other words, it can be seen as an export facility
that only exports carefully selected symbols and formulas from the
realm, without duplication or redundancy.  Note that, in practice, we
will choose for $F$ the ``usual symbols'' traditionally used for that
theory; these will also often correspond to the ``original symbols''
used in (some of) the bottom theories.

In particular, realms offer the infrastructure to satisfy the five requirements for users
of a {\UDLM} described in section~\ref{sec:motivation}.
\realmref{where} is addressed by the set of theories in the realm $R$ and by the concepts
and facts in $F$.  \realmref{minimality} is addressed by the conservative developments in
$\cC$.  \realmref{views} is addressed by the views in $\cV$ and the conservative
developments in $\cC$.  \realmref{users1} and \realmref{users2} are addressed by $F$ being
primitive and the fact that $F$ is the front of an interface to each top theory.

\begin{example}[Trivial realm]\label{ex:trivial-realm}
  Any theory $S$ in $G$ induces a simple realm $R = (G, S, \{G_S\},
  \{\mathtt{Id}_{S}\}, \{\mathtt{Id}_{S}\})$ where $G_S$ is the
  subgraph $G$ consisting of $S$ alone and $\mathtt{Id}_{S}$ is the
  identity view on $S$.  Thus $S$ serves as the top and bottom
  theories of the trivial conservative development of $S$, as well as
  the face of $R$.
\end{example}

\begin{example}[Initial realm]\label{ex:initial-realm}
  For any theory $T$ in $G$ we can extend $G$ with a copy $F_T$ of $T$ and a conservative
  interface $F_T\nmmtar{mview}{\iota}T$, where $\iota$ maps any symbol to its copy to
  obtain a theory graph $G'$. We call $R^T_G\defeq(G',F_T,\{G_T\},\{\mathtt{Id}_S\},\{\iota\})$,
  where $G_T$ is the subgraph of $G$ consisting of $T$ alone, the \defemph{initial realm} for
  $T$ in $G$.
\end{example}

\begin{wrapfigure}r{2.9cm}\vspace*{-2.2em}
\begin{tikzpicture}[scale=1.5]
  \node[thy] (t1) at (0,0) {$\top_1$};
  \node[thy] (t2) at (1.2,0) {$\top_2$};
  \node[primthy] (f1) at (0,1) {$F_1$};
  \node[primthy] (f2) at (1.2,1) {$F_2$};

  \draw[mview]  (f1) -- node[right] (i1) {$I_1$} (t1);
  \draw[mview]  (f2) -- node[right] (i2) {$I_2$} (t2);
  \draw[thy,dashed,double] (-.3,-.3) rectangle (.35,1.3);
  \draw[thy,dashed,double] (.8,-.3) rectangle (1.55,1.3);

  \draw[pviewleft]  (f1) -- node[above]{$\tilde v$} (f2);
  \draw[view]  (t1) -- node[above]{$v$} (t2);
\end{tikzpicture}\vspace*{-.8em}
\caption{Lifting}\label{fig:lifting}\vspace*{-3em}
\end{wrapfigure}
We can project any realm $R$ to its face $F$, forgetting all developmental structure. Note
that we can lift views between theories to \defemph{realm morphisms} (theory morphisms
between their faces): given two realms $R_1$ and $R_2$ with two interfaces $I_i$, fronts
$F_i$, and top theories $\top_i$, a view $\top_1\nmmtar{view}{v}\top_1$ induces a partial
view $F_1\nmmtar{pviewleft}{\tilde{v}}F_2$ on the faces, where $\tilde{v}=I_2^{-1}\circ
v\circ I_1$ (see Figure~\ref{fig:lifting}). In practice, the lifted views will almost
always be total, since we prefer to use (in $\top_i$ and $F_i$) the original symbols from
the bottom theories.

\section{Examples}\label{sec:examples}

As the development of the last few sections is fairly abstract, we
will attempt to give the reader a better feel for realms through a
selection of examples.  We develop the first one in some detail and
then give a more intuitive (and thus shorter) description of the
remaining examples.

\subsection{Groups}

It is well known that groups can alternatively be described in two ways:
\begin{boxedquote}
\noindent\emph{Definition 1}~\cite{KarMer:ftg79}. A {\bf{group$_1$}}
  is a set $G$ together with an associative binary operation
  $\circ\colon G\times G\to G$, such that there is a unit element $e$
  for $\circ$ in $G$, and all elements have inverses.

\medskip

\noindent\emph{Definition 2}~\cite{Hall:ttg59}. A {\bf{group$_2$}} is a set $G$, together
with a (not necessarily associative) binary operation $/\colon G\times G\to G$, such that
$a/a=b/b$, $a/(b/b)=a$, $(a/a)/(b/c)=c/b$, and $(a/c)/(b/c)=a/b$ for all $a,b,c\in G$.
\end{boxedquote}

\medskip

\def\lsl{{\scriptscriptstyle\kern-.2em/}}
\def\slcirc{/_{\kern-.2em\circ}}\def\circsl{\circ_\lsl}\def\esl{e_\lsl}
\def\invsl#1{#1^{-1}_\lsl} 

For any group$_1$ $(G,\circ)$, we can define a binary operation
$\slcirc$ by $a\slcirc b\defeq a\circ b^{-1}$ that shows that
$(G,\slcirc)$ is a group$_2$, and vice versa --- using $a\circsl
b\defeq a/\invsl{b}$ with $\invsl{b}\defeq(b/b)/b$.  Practitioners
want to use both group multiplication and division but are usually
indifferent to how and where they are introduced.

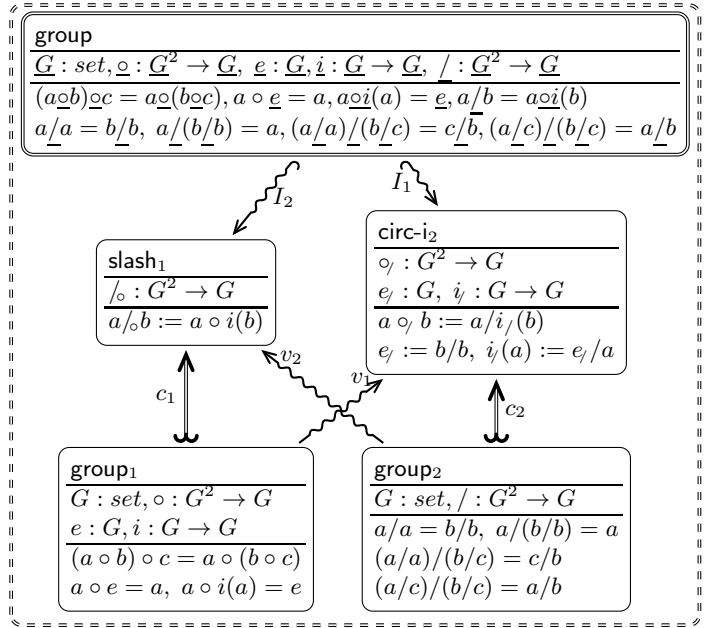
\begin{figure}[ht]\centering
\begin{tikzpicture}[scale=.75]
  \node[thy] (g1) at (0,0) {$\mmtthy{group$_1$}
           {G:set,\circ:G^2\to G\\ e:G,i:G\to G}
           {(a\circ b)\circ c=a\circ(b\circ c)\\
            a\circ e=a,\;a\circ i(a)=e}$};
  \node[thy] (g2) at (5.5,0) {$\mmtthy{group$_2$}
           {G:set,/:G^2\to G}
           {a/a=b/b,\; a/(b/b)=a\\ 
            (a/a)/(b/c)=c/b\\
            (a/c)/(b/c)=a/b}$};
  \node[thy] (s1) at (0,4.2) {$\mmtthy{slash$_1$}{\slcirc:G^2\to G}{a\slcirc b\defeq a\circ i(b)}$};
  \node[thy] (ci2) at (5.5,4.2) {$\mmtthy{circ-i$_2$}
            {\circsl:G^2\to G\\\esl:G,\;i_\lsl:G\to G}
            {a\circsl b\defeq a/i_/(b)\\\esl\defeq b/b,\;i_\lsl(a)\defeq\esl/a}$};
  \draw[conservative] (g1) -- node[left]{$c_1$} (s1);
  \draw[conservative] (g2) -- node[right]{$c_2$} (ci2);
  \draw[view] (g1) -- node[above,near end] {$v_1$} (ci2);
  \draw[view] (g2) -- node[above,near end] {$v_2$} (s1);

  \node[primthy] (g) at (3,7.9) {$\mmtthy{group}
           {\ul{G}:set,\ul\circ:\ul{G}^2\to\ul{G},\;\ul{e}:\ul{G},\ul{i}:\ul{G}\to\ul{G},\;\ul{/}:\ul{G}^2\to\ul{G}}
           {(a\ul\circ b)\ul\circ c=a\ul\circ(b\ul\circ c),
             a\circ \ul{e}=a,
             a\ul\circ \ul{i}(a)=\ul{e},
             a\ul/b = a \ul\circ \ul{i}(b)\\
             a\ul{/}a=b\ul/b,\; a\ul/(b\ul/b)=a,
             (a\ul/a)\ul/(b\ul/c)=c\ul/b,
             (a\ul/c)\ul/(b\ul/c)=a\ul/b}$};
         \draw[mviewleft]  (g) -- node[left]{$I_1$} (ci2);
  \draw[mview] (g) -- node[right]{$I_2$} (s1);
  \draw[thy,dashed,double] (-3.1,-1.7) rectangle (9.1,9.3);
\end{tikzpicture}
\caption{A Realm of Groups with Face \textsf{group}}\label{fig:groups}
\end{figure}

In Figure~\ref{fig:groups}, we have assembled this situation into a two-pillar realm with
face \textsf{group}. The two bottom theories \textsf{group}$_1$ and \textsf{group}$_2$ are
equivalent via the views $v_1$ and $v_2$ and the back views of $c_1$ and $c_1$,
respectively.\footnote{This is a very common situation; the base theories differ mainly in
  which symbols are considered primitive, and the conservative developments mainly
  introduce definitions for the remaining ones.} Note that the views
$v_1=\circ\mapsto\circsl,e\mapsto\esl,i\mapsto i_\lsl$ and $v_2=/\mapsto\slcirc$ carry
proof obligations that show that the newly defined extensions \textsf{slash}$_1$ and
\textsf{circ-i}$_2$ behave as expected by the \textsf{group}$_{3-i}$. The face
\textsf{group} contains ``new'' symbols, for which we use underlined symbols, to
distinguish them from the ones in the pillars of the realm. The interface views $I_i$ pick
out the respective ``original operators'' $/$, $\circ$, and $i$, together with the
corresponding axioms (and any theorems that may have been proven along the way). Here we
have
\[ I_1 =\ul\circ\mapsto\circ,\ul{e}\mapsto e,\ul{i}\mapsto i,\ul/\mapsto
\slcirc\quad\text{and}\quad I_2 =\ul\circ\mapsto\circsl,\ul{e}\mapsto\esl,\ul{i}\mapsto
i_\lsl,\ul/\mapsto/
\]
In particular, it is very natural to require that the interfaces of a
realm are conservative since they have access to all symbols in the
body of the realm.

\subsection{Natural Number Arithmetic}

A realm \nat of natural number arithmetic would naturally contain
conservative developments of several different axiomatizations of the
natural numbers with the usual arithmetic operations.  One
conservative development would certainly start with Peano's
axiomatization of the natural numbers~\cite{Peano89}.  The base theory
would contain the symbols 0 and $S$ (the successor function) and the
(second-order) Peano axioms.  The conservative development would
include recursive definitions of $+$ (addition) and $\ast$
(multiplication).  This development is particularly useful as it
makes the proofs of many properties of the natural numbers simple.

Another kind of conservative development would start with a construction of the natural
numbers using machinery available in the underlying logic.  There are many such
constructions.  Some examples are finite von Neumann ordinals constructed from sets,
Church numerals constructed from lambda expressions, strings of bits, and various
bijective numeration schemes.  These constructions define representations of the natural
numbers that are semantically equivalent but far from equivalent with respect to
computational complexity.  It is worth singling out the \emph{sequences of machine-sized
  words} representation, which tends to be the most efficient.

The face $F_{\nat}$ of \nat would be restricted to the most basic concepts and facts about
natural number arithmetic.  These would naturally include symbols for all the natural
numbers (i.e., natural number numerals) and all the true equations of the form $n_1 + n_2
= n_3$ and $n_1 \ast n_2 = n_3$.  Thus $F_{\nat}$ would contain an infinite number of
symbols and facts.  An implementation of $F_{\nat}$ would require an efficient means to
represent and compute with natural number numerals.  \emph{Biform theories}~\cite{Farmer:btc07} 
would be best suited for such a task.

This realm \nat that we have described is a multi-pillar presentation
of the mathematical topic of natural number arithmetic.  It can be
used by developers as a module with which to build more complex
theories and by students who are interested in understanding what are
the basic concepts and facts of this topic and how they are derived
from first principles.  A realm like \nat that contains several
pillars and a face of basic concepts and facts is called a
\defemph{foundational realm}.  Used for building new theories and for
study, a foundational realm would not be expected to change much over
time.

Since the mathematical theory of natural number arithmetic is
exceedingly rich, there are a great many concepts and facts about
natural numbers that could be of use to practitioners.  For developers
and students, the usefulness of \nat would be greatly reduced if there
was an attempt to include all of these concepts and facts in
$F_{\nat}$.  It would be much better for practitioners --- who are
primarily interested in applications --- to create another
single-pillar realm \natprime of natural number arithmetic whose face
would contain all the useful concepts and facts about natural numbers
that have been derived someplace in the theory graph.  A realm of this
type is called a \defemph{high-level realm}.  Used for applications, a
high-level realm would be continuously updated as new concepts and
facts are discovered.  It would be an implementation of the idea of a
\emph{high-level theory} discussed in~\cite{CarFar:hlt08}.

\subsection{Real Numbers}

The theory of the real numbers covers the algebraic and topological structure of the real
numbers.  Rich in concepts and facts, it is one of the most important theories in all of
mathematics.  It is important to developers since real numbers are needed in most
mathematical developments.  It is important to students since it includes many of the most
important ideas of mathematics.  It is important to practitioners since most mathematical
problems involve the real numbers in some way.

A realm \reals of the real numbers could be used to consolidate and organize all the
knowledge about the real numbers that resides in a \UDLM.  \reals would have a structure
similar to the realm of natural number arithmetic.  It would contain two kinds of
conservative developments.  The first kind are axiomatizations of a complete ordered field
-- all complete ordered fields are isomorphic.  The second kind are constructions of the
real numbers, of which there are many.  Some examples are Dedekind cuts in the field of
rational numbers, Cauchy sequences of rational numbers, infinite decimal expansions, the
quotient of the finite hyperrationals by the infinitesimal hyperrationals, and as a
substructure of the surreal numbers.  It is worth remarking that most of these
constructions leverage \nat, so that constructing realms is also a modular process.

The realm \reals would be a foundational realm like \nat for developers and students.
There should also be a high-level realm \realsprime like \natprime for practitioners.  The
face of \reals would only contain the basic concepts and facts about the real numbers,
while the face of \realsprime would contain all the useful concepts and facts about the
real numbers that have been derived someplace in the theory graph.  The prominent role of
the real numbers would mean that \reals or \realsprime would be the basis of many of the
more sophisticated theories in a \UDLM.

\subsection{Monads}

Category theorists and (advanced) Haskell programmers are familiar
with the expressive power of monads.  Most know that there are in fact
two equivalent presentations of the theory of monads, one using a
multiplication operation $\mu$ (called \texttt{join} in Haskell) and
unit $\eta$ (\texttt{return}), the other using \emph{Kleisli triples}
with a lifting operation $-^{*}$ (called \texttt{bind} or \verb|>>=|
in Haskell).  From there, one can define a large list of generic
combinators that work for any monad.

These two presentations are equivalent, and are again similar in
flavor to the previous ones: one is more convenient for proofs, the
other for computational purposes.  Again, these basic theories tend to
be followed by a substantial tower of \emph{conservative extensions}.
In other words, Haskell's \texttt{Control.Monad} should really be seen
as the \emph{face} of a realm of monads.

\subsection{Modal Logic \sfour}

The modal logic \sfour has a large number of equivalent presentations
--- John Halleck \cite{HalleckS4} lists $28$ of them.  This gives
developers significant flexibility when using views (aka requirement
\realmref{views}) to establish that a structure can interpret \sfour.
And, of course, \sfour supports rather significant conservative
extensions and applications of it are found in a variety of places.

\subsection{Models of Computation}

The \emph{Chomsky hierarchy} of regular, context-free,
context-sensitive and recursively enumerable languages offer names for
(the face of) four more, nested, realms.  As is well known, each of
the above languages contains many different formalisms which are
nevertheless equivalent.

Inside the recursively enumerable languages (for example), we would
have the pillars of Turing machines, Register Machines, the Lambda
Calculus, certain automata, etc, as alternatives.  It is difficult to
design a suitable face theory for this realm, as the syntax of any
high-level programming language could serve; given the heated
debates around what language is ``best'', this is one realm whose face
may not settle for a long time.

\section{The Realm Idea}

The examples of the previous section show the advantages of realms as consolidated
structures: a realm hides cumbersome details, while still allowing access to the details
for those (such as developers) who must deal with them.  Realms thus deal with two
structural tensions in the design of theory graphs that formalize a mathematical domain:

\paragraph{Foundational realms} can in many ways be understood as the formalization of the
ideas of \emph{information hiding} and \emph{modules} coming from software engineering.
The face of a realm corresponds to an interface; its secrets, i.e., what it hides, is the
actual conservative development of the theory; and its representation details correspond
to an axiomatization.  Of course, to get substitutivity, we need to ensure equivalence.
In an ad hoc manner, Haskell's type classes, ML's modules and functors, Scala's traits,
Isabelle's locales (etc) all capture certain aspects of realms.  However, the lack of good
support for \emph{views} really hampers the use of these proto-realms as a modular
development mechanism. 

\paragraph{High-level realms} give practitioners high-level collections of useful symbols
and formulae that function like a tool-chest for applications based on the tiny theories
developers use as a fine-grained model of dependencies, symbol visibilities, and
consistency. For them theories should be static over time, depicting a completed axiomatic
development of a mathematical topic. This gives a persistent base (and rigid designators)
to develop against. But this means that conservative extensions (like definitions and
theorems) need new theories, leading to a severe pollution of the theory namespace.
Practitioners, on the other hand, would naturally prefer dynamic theories that
continuously grow as new concepts and facts are introduced (another kind of rigid
designator).

\paragraph{The contribution of realms} is an overlay structure that can implement
information hiding, and mediates between dynamic high-level theories and an
underlying, static theory graph.  So users can have their cake and eat it.

\section{Representing and Growing Realms in a \UDLM}

Our work on OMDoc/MMT~\cite{MMTSVN:on,KohRabZho:tmlmrsca10:biblatex} and the
MathScheme~\cite{CarFarCon:mpd11} systems have given us a decent intuition (or so we
feel) regarding the services that a theory-graph based system should provide.  We now
extent this to realms.

\paragraph{Marking Up Realms.} If we look at the definition of a realm, we see that the
body components are already present in the theory graph given by the existing axiomatic
developments. Thus, given a theory graph $G$, we can add a realm $R$ by just tagging a
subgraph of $G$ and adding a set of interfaces with their common front (the face of $R$);
all of these are regular components of theory graphs, so we only need to extend the theory
graph data structures (and representation languages) by a ``realm tagging''
functionality. This also shows us that the concept of realms is conservative over theory
graphs.

One can easily envision two methods of syntactically identifying realms: globally via a
theory-level ``realm declaration'' which specifies the five components from
Definition~\ref{def:realm}, or locally by extending theory and view declarations with a
field that specifies the realm (or realms) it participates in. Given the little theory
approach, we tend to use theory extensions and view declarations when the local context
is clear (for example, within a single ``file''), and the more global approach when
drawing from a wider context.  This appears to be a good syntactic compromise.

An implementation will have to check the internal constraints from
Definition~\ref{def:realm}, in particular, that interfaces are total. But the idea of
simply ``discovering'' realms that occur in the wild is a bit optimistic. From our case
studies, we expect that realms have to be engineered purposefully: they are grown from a
seed, and grow over time by coordinated (and system-supported) additions of theories and
views.

\subsection{Supporting the Life Cycle of Realms}

\begin{sloppypar} 
We postulate that three realm-level operations will be needed in practice:
\begin{inparaenum}[\em i\rm)]
\item realms are initialized by designating chosen theories as initial realms, which
\item can be extended by adding conservative extensions, and
\item proper realms are created by merging existing realms.
\end{inparaenum}
These three operations were sufficient to explain the complex realms in our case
studies. We will now discuss them in more detail.
\end{sloppypar}

\begin{wrapfigure}r{2.5cm}\footnotesize\vspace*{-3em}
  \begin{tikzpicture}[xscale=.5,yscale=1.3]
    \node[thy] (b) at (0,0) {$\bot$};
    \node[thy] (s) at (0,1) {$S$};
    \draw[conservdev] (b) -- (s);
    \node[thy] (t) at (0,2) {$\top$};
    \draw[conservdev] (s) -- (t);
    \node[thy,double] (f) at (0,3) {$F$};
    \draw[mview] (f) -- node[left]{$I$} (t);

    \node at (1.2,2) {$\leadsto$};

    \node[thy] (nb) at (3,0) {$\bot$};
    \node[thy] (ns) at (3,1) {$S$};
    \draw[conservdev] (nb) -- (ns);
    \node[thy] (sp) at (2.3,2) {$S'$};
    \draw[conservative] (ns) -- (sp);
    \node[thy] (nt) at (3.7,2) {$\top$};
    \draw[conservdev] (ns) -- (nt);
    \node[thy] (tp) at (3,3) {$\top'$};
    \draw[conservdev] (sp) -- (tp);
    \draw[conservative] (nt) -- (tp);
    \node[thy,double] (nf) at (3,4) {$F'$};
    \draw[mview] (nf) -- node[right]{$I'$} (tp);
  \end{tikzpicture}
  \caption*{realm extension}\vspace*{-3em}
\end{wrapfigure}
\paragraph{Initializing Realms.} Given a theory graph $G$, we add any realm (e.g., the
initial realm $R^T_G$ for a theory $T$ in $G$; see Example~\ref{ex:initial-realm}) as a
starting point of development.
\paragraph{Extending Realms by (internal) conservative extensions.}
Given a realm $R\defeq(G,F,\cC,\cV,\cI)$, a top theory $\top$ of some $C\in\cC$, and an
interface $I$ for $\top$, then we can extend $R$ by:
\begin{compactenum}[\em i\rm)]
\item adding a conservative extension $S\mmtar{conservative}S'$ by declaration $c$ and a
  commensurate extension $\top\mmtar{conservative}\top'$ to $C$ and
\item (optionally) adding a declaration $\ul{c}$ to $F$, giving a new face $F'$, and
  extending $I$ so that $I':=I,\underline{c}\mapsto c$. If we do -- e.g., for a high-level
  view -- we have to apply \emph{i}) to each of the pillars of $R$, so that all their
  interfaces are total; the diagram shows the situation for a simple realm.
\end{compactenum}

In particular, an implementation of high-level realms must provide a registration
functionality for conservative extensions in $\cC$ that keeps the interface(s) consistent
by ensuring new names appear in the face of the realm. Note that this extension operation
does not change the number of pillars of a realm, in particular, if realms are started by
initial realms, they will only be extended to simple realms. It is predominantly used for
high-level realms.

The next operation merges two realms if they are mutually interpretable. This operation is
mainly used to build up foundational views, the construction makes sure that all symbols
in the face are interpreted in all the pillars. 
  
\begin{figure}[ht]\centering
  \begin{tikzpicture}[yscale=1.5,xscale=1.2]
    \node[thy] (b1) at (-4,0) {$\bot_1$};
    \node[thy] (b2) at (-2.5,0) {$\bot_2$};
    \draw[view] (b1) to[out=15,in=165] node[above] {$v$} (b2);
    \draw[view] (b2) to[out=195,in=-15] node[below]{$w$}(b1);

    \node[thy] (t1) at (-4,1) {$\top_1$};
    \node[thy] (t2) at (-2.5,1) {$\top_2$};
    \node[thy] (t2) at (-2.5,1) {$\top_2$};
    \draw[conservdev] (b1) -- (t1);
    \draw[conservdev] (b2) -- (t2);
    \node[thy,double] (f1) at (-4,2) {$F_1$};
    \node[thy,double] (f2) at (-2.5,2) {$F_2$};
    \draw[mview] (f1) -- node[left]{$I_1$} (t1);
    \draw[mview] (f2) -- node[left]{$I_2$}  (t2);

    \node at (-1.5,1) {$\leadsto$};

    \node[thy] (b1) at (0,0) {$\bot_1$};
    \node[thy] (b2) at (3,0) {$\bot_2$};
    \draw[view] (b1) to[out=10,in=170] node[above,near start] {$v$} (b2);
    \draw[view] (b2) to[out=190,in=-10] node[below,near start]{$w$}(b1);

    \node[thy] (t1) at (-.8,1) {$\top_1$};
    \node[thy] (pt2) at (.8,1) {$w(\top_2)$};
    \node[thy] (pt1) at (2.2,1) {$v(\top_1)$};
    \node[thy] (t2) at (3.8,1) {$\top_2$};
    \draw[conservdev] (b1) -- (t1);
    \draw[conservdev] (b2) -- (t2);
    \draw[conservdev] (b1) -- (pt2);
    \draw[conservdev] (b2) -- (pt1);
    \draw[view] (t1) to[out=20,in=160] node[above,near end] {$v$} (pt1);
    \draw[view] (t2) to[out=200,in=-20] node[below,near end]{$w$}(pt2);

    \node[thy] (tp1) at (0,2) {$\top_1\cup w(\top_2)$};
    \node[thy] (tp2) at (3,2) {$\top_2\cup v(\top_1)$};
    \draw[view] (tp1) to[out=10,in=170] node[above] {$v$} (tp2);
    \draw[view] (tp2) to[out=190,in=-10] node[below,near start]{$w$}(tp1);
    \draw[conservdev] (t1) -- (tp1);
    \draw[conservdev] (t2) -- (tp2);
    \draw[conservdev] (pt2) -- (tp1);
    \draw[conservdev] (pt1) -- (tp2);
    \node[thy,double] (f) at (1.5,2.7) {$F_1\cup F_2$};
    \draw[mview] (f) -- node[above,near end]{$I_1^{+w}$}  (tp1);
    \draw[mviewleft] (f) -- node[above,near end]{$I_2^{+v}$}  (tp2);
  \end{tikzpicture}
  \caption{Union Realm}\label{fig:union-realm}
  \vspace*{-3em}
\end{figure}
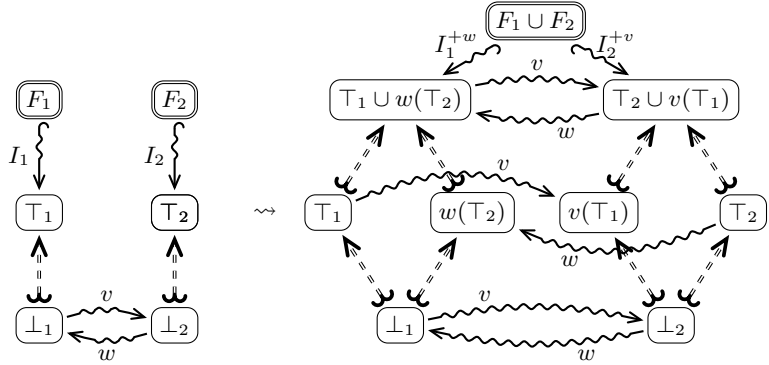

\paragraph{Merging Realms along Views.} Given two realms
$R_1\defeq(G,F_1,\cC_1,\cV_1,\cI_1)$ and $R_2\defeq(G,F_2,\cC_2,\cV_2,\cI_2)$, and views
$\bot_1\nmmtar{view}{v}\bot_2$ and $\bot'_2\nmmtar{view}{w}\bot'_1$, where $\bot_i$ and
$\bot'_i$ are (arbitrary) bottom theories in $\cC_i$, then we can define the union realm
$R_1\cup^v_w R_2$ along $v$ and $w$ as $(G,F_1\cup F_2,\cC_1^{+w}\cup
\cC_2^{+v},\cV_1\cup\cV_2\cup\{v,w\},\cI_1^{+w}\cup\cI_2^{+v})$.  Figure~\ref{fig:union-realm}
shows the situation for two simple realms.  Generally, we define that:
\begin{compactenum}[\em i\rm)]
\item $\cC_1^{+w}$ is the set of conservative developments $\{C^{+w}~|~C\in\cC_1\}$, where
  $C^{+w}$ is $C$ extended by a copy\footnote{A copy of a development (sub)graph $H$ along a
    view $v$ is an isomorphic graph $H'$, where for any theory $S$ in $H$, 
    $S'$ in $H'$ consists of the declarations
    $c:v(\tau)=v(\delta)$, for all $c:\tau=\delta$ in $S$. This construction gives us
    a view $S\nmmtar{view}{v}S'$.} of the 
  development of $\bot'_2$ to $\top'_2$ along $w$, itself extended to
  $\top\cup w(\top'_2)$. $\cC_2^{+v}$ is
  defined analogously. In Figure~\ref{fig:union-realm}, $\cC_2^{+v}$ and $\cC_1^{+w}$ are
  the two diamonds on the left and right.
\item $F_1\cup F_2\nmmtar{mview}{I_1^{+w}}\top_1\cup w(\top_2)$ is $I_1\cup w\circ I_2$
  and $F_1\cup F_2\nmmtar{mview}{I_2^{+v}}\top_2\cup v(\top_1)$ is $I_2\cup v\circ I_1$.
\end{compactenum}
An implementation of this construction would take great care to merge corresponding
symbols in the two faces to minimize the union. Moreover, the copying operation can be
optimized to only copy over those conservative extensions that are mentioned in the
interface extension.

\subsection{Modular Realms}
Note that the extension and merging operations highlight an internal
invariant of realms that may not have been obvious until now: All
pillars of a realm must interpret the full vocabulary of the face to
admit total interfaces. This duplication can become quite tedious in
practice. Therefore it is good practice to modularize realms in the
spirit of a ``little realms approach''.  For instance, the groups
realm from Figure~\ref{fig:groups} could be extended by the usual
group theorems via conservative extensions in both pillars. But we can
also build a simple realm with base theory \textsf{group} and extend
that conservatively (once per theorem). Unless there are proofs that
directly profit from the particulars of the concrete formulations in
the pillars below, the modular approach is more efficient
representationally and thus more manageable.

\subsection{Interface Matters}\label{sec:ui}

As the realms are the main interaction points for mathematicians with the \UDLM, realms
must be discoverable and provide a range of convenient information retrieval methods
(after all, realms will get very large in practice). These can range from community tools
like peer reviewed periodicals (aka. academic journals) to technical means like 
intra- and cross-realm search engines (as realms are built upon theory graphs,
specialization of the \textsf{$\flat$search} engine~\cite{KohIan:ssmk12} will be a good
starting point.)

It will be very important to provide a set of interactions for the interface of a realm
that users can understand. It will be important to look up the definienda and proofs of
interface items, even though this will usually mean that we need to descend into
(conservative extensions of) one of the fronts of the interface, which employ different
languages. This needs to be transparent enough to be understandable to
users/mathematicians.

Similarly, the equivalence relation of the (tiny) theories that make up the realm should
be made transparent and easy to browse for the user.

\section{Conclusion}\label{sec:concl}
We have presented an extension of the theory graph approach to
representing mathematical knowledge. Realms address the mismatch
between the successful practice of the little/tiny theory approach
natural for developing theory graphs and the high-level theories most
useful for practitioners utilizing such mathematical knowledge
representations. We have proposed a formal definition for
realms that is conservative over theory graphs and
shown its adequacy by applying it to examples from various areas of
mathematics and computation.

As a step towards an implementation we have investigated a set of realm-level operations
that can serve as a basis for system support of realm management.  The next step in our
investigation will be to realize and test such support in the
OMDoc/MMT~\cite{MMTSVN:on,KohRabZho:tmlmrsca10:biblatex} and the
MathScheme~\cite{CarFarCon:mpd11} systems, fully develop the examples sketched in this
paper, and test the interactions on developers, students, and practitioners (see
section~\ref{sec:motivation}).

\begin{sloppypar}
\printbibliography
\end{sloppypar}
\end{document}

